\documentclass[aps,twocolumn,showpacs]{revtex4}

\usepackage{epsf}
\usepackage{bm}

\begin{document}

\title{Asymptotic exchange coupling of quasi-one-dimensional excitons in carbon nanotubes}

\author{I.V.~Bondarev}\email[Corresponding author.
E-mail: ]{ibondarev@nccu.edu}\affiliation{Physics Department,
North Carolina Central University, 1801 Fayetteville Str, Durham,
NC 27707, USA}

\begin{abstract}
An analytical expression is obtained for the biexciton binding
energy as a function of the inter-exciton distance and binding
energy of constituent quasi-one-dimensional excitons in carbon
nano\-tubes. This allows one to trace biexciton energy variation
and relevant non-linear absorption under external conditions
whereby the exciton binding energy varies. The non-linear
absorption lineshapes calculated exhibit characteristic asymmetric
(Rabi) splitting as the exciton energy is tuned to the nearest
interband plasmon resonance. These results are useful for tunable
optoelectronic device applications of optically excited
semiconducting carbon nanotubes, including the strong excitation
regime with optical non-linearities.
\end{abstract}

\maketitle


Single-walled carbon nanotubes (CNs) --- graphene sheets rolled-up
into cylinders of $\sim\!1\!-\!10$~nm in diameter and
$\sim\!1\mu$m up to $\sim\!1$~cm in length~\cite{Saito,Huang} ---
are shown to be very useful as miniaturized electromechanical and
chemical devices~\cite{Baughman}, scanning probe
devices~\cite{Popescu}, and nanomaterials for macroscopic
composites~\cite{cncomposites}. The area of their potential
applications was recently expanded to
nanophotonics~\cite{Bond10jctn09oc,Xia08} after the demonstration
of controllable single-atom incapsulation into single-walled
CNs~\cite{Jeong}, and even to quantum cryptography since the
experimental evidence was reported for quantum correlations in the
photoluminescence spectra of individual nanotubes~\cite{Imamoglu}.

The true potential of CN-based optoelectronic device applications
lies in the ability to tune their properties in a precisely
controllable way. In particular, optical properties of
semiconducting CNs originate from excitons, and may be tuned by
either electrostatic doping~\cite{Spataru10}, or via the quantum
confined Stark effect (QCSE) by means of an electrostatic field
applied perpendicular to the CN axis~\cite{Bond09PRB}. In both
cases the exciton properties are mediated by collective plasmon
excitations in CNs~\cite{cnplasmons}. In the case of the
perpendicularly applied electrostatic field, in particular, we
have shown recently~\cite{Bond09PRB} that the QCSE allows one to
control the exciton-interband-plasmon coupling in individual
undoped CNs and their (linear) optical absorption properties,
accordingly.

Here, I extend our studies to the strong (non-linear) excitation
regime whereby photogenerated biexcitonic states may be formed in
CNs (observed recently in single-walled CNs by the femtosecond
transient absorption spectroscopy
technique~\cite{Papanikolas}).~An analytical (universal)
expression is obtained for the biexciton binding energy as a
function of the inter-exciton distance and the binding energy of
constituent excitons. The formula~is consistent with the numerical
results reported earlier~\cite{Pedersen05,Molinari07}, and is
advantageous in that it allows one to trace biexciton energy
variation and relevant non-linear absorption, accordingly, as the
exciton energy is tuned to the nearest interband plasmon resonance
by means of the QCSE. The non-linear absorption lineshapes are
calculated close to the first interband plasmon resonance for the
semiconducting (11,0) CN (chosen as an example) under resonant
pumping conditions~\cite{Mukamel}. They exhibit the characteristic
asymmetric splitting behavior similar to that reported for the
linear absorption regime~\cite{Bond09PRB}. This effect could help
identify the presence and study the properties of biexcitons in
individual single-walled CNs, which is not an easy task under
non-linear excitation because of the strong competing
exciton-exciton annihilation process~\cite{THeinz,Valkunas,Kono}.

\begin{figure}[t]
\epsfxsize=7.5cm\centering{\epsfbox{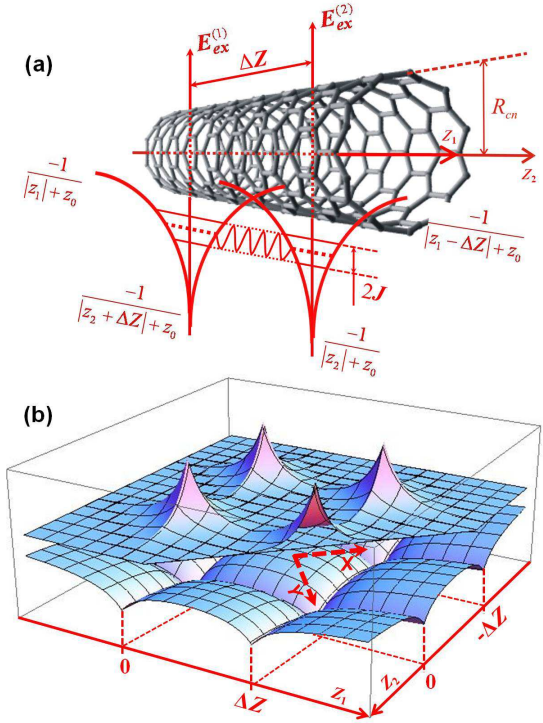}}\caption{(Color
online) (a)~Schematic of the exchange coupling of two ground-state
1D excitons to form a~biexcitonic~state (arb.~units). Two
collinear axes, $z_1$ and $z_2$, represent independent relative
electron-hole motions in the 1st and 2nd exciton, have their
origins shifted by $\Delta Z$, the inter-exciton center-of-mass
separation. (b)~The coupling occurs in the configuration space of
the two independent longitudinal relative electron-hole motion
coordinates, $z_1$ and $z_2$, of each of the excitons, due to the
tunneling of the system through the potential barriers formed by
the two single-exciton cusp-type potentials [bottom, also in (a)],
between equivalent states represented by the isolated two-exciton
wave functions shown on the top.}\label{fig1}
\end{figure}

The binding energy of the biexciton in a small-diameter
($\sim\!1\,$nm) CN can be evaluated by the method pioneered by
Landau~\cite{LandauKM}, Gor'kov and Pitaevski~\cite{Pitaevski},
Holstein and Herring~\cite{Herring} --- from the analysis of the
asymptotic exchange coupling by perturbation on the configuration
space wave function of the two ground-state one-dimensional (1D)
excitons. Using the cylindrical coordinate system with the
\emph{z}-axis along the CN axis and separating out circumferential
and longitudinal degrees of freedom of each of the excitons by
transforming their longitudinal motion into their respective
center-of-mass coordinates~\cite{Bond09PRB,Ogawa}, one arrives at
the biexciton Hamiltonian of the form [see Fig.~\ref{fig1}~(a)]
\begin{eqnarray}
\hat{H}(z_1,z_2,\Delta
Z)=-\frac{\partial^2}{\partial\,\!z_{1}^2}-\frac{\partial^2}{\partial\,\!z_{2}^2}\hskip2cm\label{biexcham}\\
-\frac{1}{|z_{1}|\!+\!z_0}-\!\frac{1}{|z_{1}\!-\!\Delta Z|\!+\!z_0}
-\!\frac{1}{|z_{2}|\!+\!z_0}-\!\frac{1}{|z_{2}\!+\!\Delta Z|\!+\!z_0}\nonumber\\
-\frac{2}{|(z_1+z_2)/2+\Delta Z|\!+\!z_0}-\frac{2}{|(z_1+z_2)/2-\Delta Z|\!+\!z_0}\;\;\nonumber\\
+\frac{2}{|(z_1-z_2)/2+\Delta Z|\!+\!z_0}+\frac{2}{|(z_1-z_2)/2-\Delta
Z|\!+\!z_0}\;.\nonumber
\end{eqnarray}
Here, $z_{1,2}\!=\!z_{e1,2}-z_{h1,2}$ are the relative electron-hole
motion coordinates of two 1D excitons separated by the
center-of-mass-to-center-of-mass distance $\Delta Z\!=\!Z_2-Z_1$, $z_0$ is
the cut-off parameter of the effective (cusp-type) longitudinal electron-hole Coulomb
potential.~Equal electron and hole effective masses $m_{e,h}$ are assumed~\cite{Jorio}
and "atomic units"\space are used~\cite{LandauKM,Pitaevski,Herring}, whereby
distance and energy are measured in units of the exciton Bohr
radius and Rydberg energy, $a^\ast_B$ and $Ry^\ast\!=\hbar^2/(2\mu a_B^{\ast2})$, respectively,
$\mu\,(\approx\!m_e/2)$ is the exciton reduced mass. First two lines in Eq.~(\ref{biexcham}) represent
two non-interacting 1D excitons with their individual potentials symmetrized
to account for the presence of the neighbor a distance~$\Delta Z$ away, as seen from the $z_1$-
and $z_2$-coordinate systems treated independently [Fig.~\ref{fig1}~(a)].~Last two lines are
the inter-exciton exchange Coulomb interactions ---
electron-hole and hole-hole + electron-electron, respectively~\cite{Endnote1}.
Biexciton binding energy is $E_{X\!X}=E_g-2E_X$, where $E_g$ is
the lowest eigenvalue of Eq.~(\ref{biexcham}), $E_X=-Ry^\ast/\nu_0^2$ is
the single exciton binding energy with $\nu_0$ being the
lowest-bound-state quantum number of the 1D exciton~\cite{Ogawa}. Negative $E_{X\!X}$
indicates that the biexciton is stable with respect to
dissociation into two isolated excitons.


The Hamiltonian (\ref{biexcham}) is effectively two dimensional in
the configuration space of the two \emph{independent} relative
motion coordinates, $z_1$ and $z_2$. Figure~\ref{fig1}~(b),
bottom, shows schematically the potential energy surface of the
two closely spaced non-interacting 1D excitons [second line of Eq.~(\ref{biexcham})]
in the $(z_1,z_2)$ space. The surface has four symmetrical minima [representing
isolated two-exciton states shown in Fig.~\ref{fig1}~(b), top], separated by the potential barriers
responsible for the tunnel exchange coupling between the
two-exciton states in the configuration space. The coordinate
transformation $x=(z_1-z_2-\Delta Z)/\sqrt{2}$,
$y=(z_1+z_2)/\sqrt{2}$ places the origin of the new coordinate
system into the intersection of the two tunnel channels between
the respective potential minima [Fig.~\ref{fig1}~(b)], whereby the
exchange splitting formula of
Refs.~\cite{LandauKM,Pitaevski,Herring} takes the form
\begin{equation}
E_{g,u}(\Delta Z)-2E_X=\mp J(\Delta Z), \label{Egu}
\end{equation}
where $E_{g,u}$ are the ground-state and excited-state energies
[eigenvalues of Eq.~(\ref{biexcham})] of the two coupled excitons
as functions of their center-of-mass-to-center-of-mass separation,
and
\begin{equation}
J(\Delta Z)=\frac{2}{3!}\int_{\!-\Delta Z/\!\sqrt{2}}^{\Delta
Z/\!\sqrt{2}}\!dy\left|\psi(x,y)\frac{\partial\psi(x,y)}{\partial
x}\right|_{\!x=0} \label{J}
\end{equation}
is the tunnel exchange coupling integral, where $\psi(x,y)$ is the
solution to the Schr\"{o}dinger equation with the Hamiltonian
(\ref{biexcham}) transformed to the $(x,y)$ coordinates. The
factor $2/3!$ comes from the fact that there are two equivalent
tunnel channels in the problem, mixing three equivalent
indistinguishable two-exciton states in the configuration space
[one state is given by the two minima on the $x$-axis, and two
more are represented by each of the minima on the $y$-axis ---
compare Fig.~\ref{fig1}~(a) and (b)].

The function $\psi(x,y)$ in Eq.~(\ref{J}) is sought in the form
\begin{equation}
\psi(x,y)=\psi_0(x,y)\exp[-S(x,y)]\,, \label{psixy}
\end{equation}
where $\psi_0=\nu_0^{-1}\exp[-(|z_1(x,y,\Delta Z)|+|z_2(x,y,\Delta
Z)|)/\nu_0]$ is the product of two single-exciton wave
functions~\cite{Endnote2} representing the isolated two-exciton
state centered at the minimum $z_1\!=\!z_2\!=\!0$ (or
$x\!=\!-\Delta Z/\sqrt{2}$, $y\!=\!0$) of the configuration space
potential [Fig.~\ref{fig1}~(b)], and $S(x,y)$ is a slowly varying
function to take into account the deviation of $\psi$ from
$\psi_0$ due to the tunnel exchange coupling to another equivalent
isolated two-exciton state centered at $z_1\!=\Delta Z$,
$z_2\!=\!-\Delta Z$ (or $x\!=\!\Delta Z/\sqrt{2}$, $y\!=\!0$).
Substituting Eq.~(\ref{psixy}) into the Schr\"{o}dinger equation
with the Hamiltonian (\ref{biexcham}) pre-transformed to the
$(x,y)$ coordinates, one obtains in the region of interest ($z_0$ dropped~\cite{Endnote2})
\[
\frac{\partial S}{\partial x}=\nu_0\left(\frac{1}{x+3\Delta
Z/\sqrt{2}}-\frac{1}{x-\Delta Z/\sqrt{2}}\right.
\]\vskip-0.37cm
\[
\left.+\frac{1}{y-\sqrt{2}\Delta Z}-\frac{1}{y+\sqrt{2}\Delta
Z}\right),
\]
up to negligible terms of the order of the inter-exciton~van der
Waals energy and up to second derivatives of~$S$. This equation is
to be solved with the boundary condition $S(-\Delta
Z/\sqrt{2},y)\!=\!0$ originating from the natural requirement
$\psi(-\Delta Z/\sqrt{2},y)\!=\!\psi_0(-\Delta Z/\sqrt{2},y)$, to
result in
\begin{equation}
\vspace{-0.35cm}
S(x,y)=\nu_0\!\left(\!\ln\!\left|\frac{x\!+\!3\Delta
Z/\!\sqrt{2}}{x-\Delta Z/\!\sqrt{2}}\right|+\frac{2\sqrt{2}\Delta
Z(x\!+\!\Delta Z/\!\sqrt{2})}{y^2-2\Delta Z^2}\!\right)\!.
\label{sxy}
\end{equation}

After plugging Eqs.~(\ref{sxy}) and (\ref{psixy}) into
Eq.~(\ref{J}), and retaining only the leading term of the integral
series expansion in powers of $\nu_0$ subject to $\Delta Z>1$, one
obtains
\begin{equation}
J(\Delta
Z)=\frac{2}{3\nu_0^3}\left(\frac{e}{3}\right)^{2\nu_0}\!\!\!
\Delta Z\,e^{-2\Delta Z/\nu_0}. \label{Jfin}
\end{equation}
The ground state energy $E_g$ of two coupled 1D excitons in Eq.~(\ref{Egu}) is now
seen to go through the negative minimum (biexcitonic state) as
$\Delta Z$ increases. The minimum occurs at $\Delta Z_0=\nu_0/2$, whereby the biexciton binding energy~is
$E_{X\!X}\!=\!-J(\Delta Z_0)\!=\!-(1/9\nu_0^2)(e/3)^{2\nu_0-1}$. In absolute units,
expressing $\nu_0$ in terms of $E_X$, one has
\begin{equation}
E_{X\!X}=-\frac{1}{9}\;|E_X|\left(\frac{e}{3}\right)^{2\sqrt{Ry^\ast/|E_X|}\,-\,1}\!\!\!.
\label{Exx}
\end{equation}

The energy $E_{X\!X}$ can be affected by the QCSE as $|E_X|$
decreases quadratically with the perpendicular electrostatic field
applied~\cite{Bond09PRB}. The field dependence in Eq.~(\ref{Exx})
mainly comes from the pre-exponential factor. So, $|E_{X\!X}|$
will be decreasing quadratically with the field, as well, for not
too strong perpendicular fields.~At the same time, the equilibrium
inter-exciton separation in the biexciton, $\Delta
Z_0=\nu_0/2=\!1/(2\sqrt{|E_X|}\,)$ (atomic units), will be slowly
increasing with the field, consistently with the lowering of
$|E_{X\!X}|$. In the zero field, assuming $|E_X|\sim r^{-0.6}$
($r$ is the dimensionless CN radius) as reported earlier from
variational calculations~\cite{Pedersen03}, one has
$|E_{X\!X}|\sim r^{-0.6}$ as well, which is weaker than
the~$r^{-1}$ dependence of Ref.~\cite{Pedersen05}, but agrees
qualitatively with the recent advanced Monte-Carlo
simulations~\cite{Molinari07}.~Interestingly, as $r$ goes down,
the ratio $|E_{X\!X}/E_X|$ in Eq.~(\ref{Exx}) slowly grows up
approaching the 1D limit $1/3e\approx0.12$. This tendency can also
be traced in the Monte-Carlo data of Ref.~\cite{Molinari07}.
Finally, $\Delta Z_0$ goes down with decreasing $r$, thus
explaining experimental evidence for enhanced exciton-exciton
annihilation in small diameter CNs~\cite{THeinz,Valkunas,Kono}.

\begin{figure}[t]
\epsfxsize=8.00cm\centering{\epsfbox{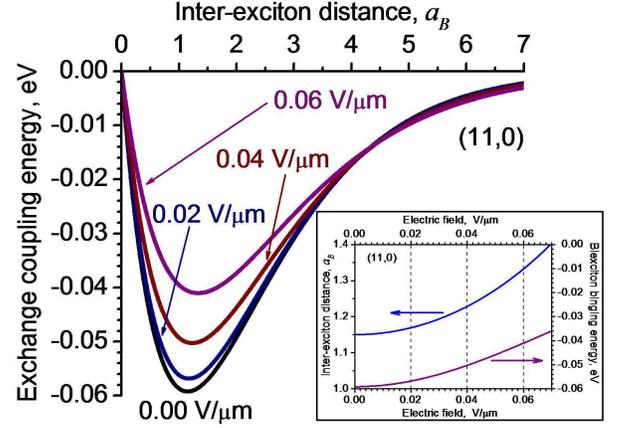}}\vskip-0.4cm\caption{(Color
online) Difference $E_g(\Delta Z)-2E_X$ for the coupled pair of
the first bright excitons in the (11,0)~CN as a function of the
center-of-mass-to-center-of-mass inter-exciton separation $\Delta
Z$ and perpendicular electrostatic field applied. Inset:~biexciton
binding energy $E_{X\!X}$ and equilibrium inter-exciton separation
$\Delta Z_0$ ($y$- and $x$-coordinates, respectively, of the
minima in the main figure) as functions of the field.}\label{fig2}
\end{figure}

Figure~\ref{fig2} shows the difference $E_g(\Delta
Z)-2E_X\!=\!\!-J(\Delta Z)$ calculated from Eqs.~(\ref{Egu}) and
(\ref{Jfin}) for a specific example of the coupled pair of the
first bright excitons in the semiconducting (11,0) CN exposed to
different perpendicular electrostatic fields. The inset shows the
field dependences of $E_{X\!X}$ [as given by Eq.~(\ref{Exx})] and
of $\Delta Z_0$. All the curves are calculated using
$Ry^\ast\!=4.02$~eV, $|E_X|=0.76$~eV, and the field dependence of
$E_X$ reported earlier in Ref.~\cite{Bond09PRB}. They exhibit
typical behaviors discussed above.

Now consider the exciton absorption lineshape under controlled
(e.g., by the QCSE) variable exciton-interband -plasmon
coupling~\cite{Bond09PRB}. In the linear (longitudinal) excitation
regime, one has for the exciton with the energy $\varepsilon$
close to a plasmon resonance the lineshape of the form
\begin{equation}
I(x)=\frac{I_{0}(\varepsilon)\,[(x-\varepsilon)^{2}+\Delta
x_p^2]}{[(x-\varepsilon)^{2}-X^{2}/4]^{2}+(x-\varepsilon)^{2}(\Delta
x_p^2+\Delta\varepsilon^2)}\,, \label{Ixfin}
\end{equation}
where $I_{0}\!=\!\Gamma(\varepsilon)/2\pi$, $\Gamma$ is the
spontaneous decay rate into plasmons, $X\!=\!\sqrt{4\pi\Delta
x_p\,I_{0}}$, $\,\Delta x_p$ is the half-width-at-half-maximum of
the plasmon resonance with the energy $x_p\,$, and
$\Delta\varepsilon$ is an additional exciton energy broadening
(normally attributed to the exciton-phonon scattering with the
relaxation time $\tau_{ph}$). All quantities in Eq.~(\ref{Ixfin})
are dimensionless, i.e. normalized to $2\gamma_0$, where
$\gamma_0\!=\!2.7$~eV is the C-C overlap integral, and the
condition $\varepsilon\!\sim\!x_p$ is assumed to hold.

The non-linear optical susceptibility is proportional to the
linear optical response function under resonant pumping
conditions~\cite{Mukamel}. This allows one to use
Eq.~(\ref{Ixfin}) to study the non-linear excitation regime with
the photoinduced biexciton formation as the exciton energy is
tuned to the nearest interband plasmon resonance. The third-order
longitudinal CN susceptibility is then of the
form~\cite{Mukamel,Pedersen05}
\begin{equation}
\chi^{(3)}(x)=\chi_0\;I(x)\left[\frac{1}{x-\varepsilon+i(\Gamma/2+\Delta\varepsilon)}\right.
\label{chi3x}
\end{equation}\vskip-0.4cm
\[
\left.-\frac{1}{x-(\varepsilon-|\varepsilon^{X\!X}|)+i(\Gamma/2+\Delta\varepsilon)}\right],
\]
where $\varepsilon^{X\!X}\!=E_{X\!X}/2\gamma_0$ is the
dimensionless binding energy of the biexciton composed of two
(ground-internal-state) excitons, and $\chi_0$ is the
frequency-independent constant. The first and second terms in the
brackets represent bleaching due to the depopulation of the ground
state and photoinduced absorption due to exciton-to-biexciton
transitions, respectively.

\begin{figure}[t]
\epsfxsize=8.65cm\centering{\epsfbox{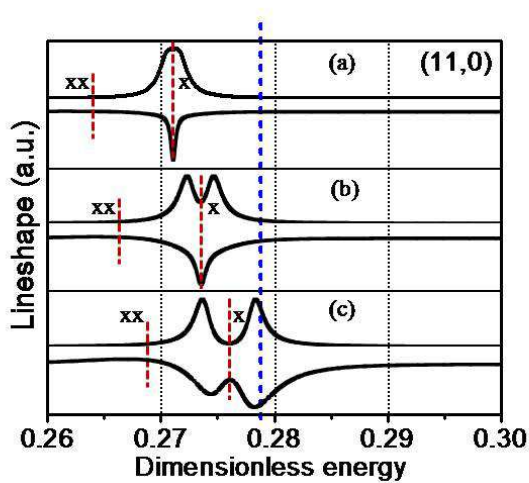}}\vskip-3cm\caption{(Color
online) [(a), (b), and (c)] Linear (top) and non-linear (bottom)
response functions as given by Eq.~(\ref{Ixfin}) and by the
imaginary part of Eq.~(\ref{chi3x}), respectively, for the first
bright exciton in the (11,0) CN as the exciton energy is tuned to
the nearest interband plasmon resonance (vertical dashed line).
Vertical lines marked as X and XX show the exciton energy and
biexciton binding energy, respectively. Dimensionless energy is
defined as $[Energy]/2\gamma_0$, where $\gamma_0=2.7$~eV is the
C-C overlap integral.}\label{fig3}
\end{figure}

Figure~\ref{fig3} compares the linear response lineshape
(\ref{Ixfin}) with the imaginary part of Eq.~(\ref{chi3x})
representing the non-linear optical response function under
resonant pumping, both calculated for the first bright
(ground-internal-state) exciton in the (11,0) CN, as its energy is
tuned to the nearest interband plasmon resonance (vertical dashed
line in Fig.~\ref{fig3})~\cite{Bond09PRB}. In this calculation,
$E_{X\!X}$ was taken to be $-0.059$~eV as given by Eq.~(\ref{Exx})
in the zero field~\cite{Endnote3}. (Weak field dependence of
$E_{X\!X}$ does not play an essential role here as
$|E_{X\!X}|\ll|E_X|=0.76$~eV regardless of the field strength.)
The phonon relaxation time $\tau_{ph}\!=\!30$~fs was used as
reported in Ref.~\cite{Perebeinos05}, since this is the shortest
one out of possible exciton relaxation processes, including
exciton-exciton annihilation
($\tau_{ee}\!\sim\!1$~ps~\cite{THeinz}). Rabi splitting
$\sim\!0.1$~eV is seen both in the linear and in non-linear
excitation regime, indicating the strong exciton-plasmon coupling
both in the single-exciton and in biexciton states, almost
unaffected by the phonon relaxation.

This effect can be used in tunable optoelectronic device
applications of small-diameter semiconducting CNs in areas such as
nanophotonics, nanoplasmonics, and cavity quantum electrodynamics,
including the strong excitation regime with optical
non-linearities. In the latter case, the experimental observation
of the non-linear absorption line splitting predicted here would
help identify the presence and study the properties of biexcitonic
states (including biexcitons formed by excitons of different
subbands~\cite{Papanikolas}) in individual single-walled CNs, due
to the fact that when tuned close to a plasmon resonance the
exciton relaxes into plasmons at a rate much greater than
$\tau_{ph}^{-1}\;(\,\gg\!\tau_{ee}^{-1})$, totally ruling out the
role of the competing exciton-exciton annihilation process.

Support from NSF (ECCS-1045661 \& HRD-0833184), NASA (NNX09AV07A),
and ARO (W911NF-10-1-0105) is gratefully acknowledged.


\begin{thebibliography}{[99]}
\bibitem{Endnote1}The intra-exciton motion is assumed to be faster than
the inter-exciton center-of-mass relative motion, thus justifying
the adiabatic approximation used in Eq.~(\ref{biexcham}).
\bibitem{Endnote2}This is an approximate solution to the Shr\"{o}dinger equation with the Hamiltonin given
by the first two lines in Eq.~(\ref{biexcham}), where the cut-off
parameter $z_0$ is neglected~\cite{Ogawa}. This approximation
greatly simplifies problem solving, while still remaining adequate
as only the long-distance tail of $\psi_0$ is important for the
tunnel exchange coupling.
\end{thebibliography}
\end{document}